\documentclass[a4paper]{jpconf}
\pdfoutput=1
\usepackage{graphicx}
\begin{document}
\title{Predicting dataset popularity for the CMS experiment}
\author{V. Kuznetsov$^1$, T. Li$^1$, L. Giommi$^2$, D. Bonacorsi$^2$, T. Wildish$^3$}
\address{$^1$ Cornell University, Ithaca, NY 14850, USA}
\address{$^2$ University of Bologna, INFN-Bologna, Italy}
\address{$^3$ Princeton University, New Jersey, NJ 08544, USA}
\ead{vkuznet@gmail.com}

\begin{abstract}
The CMS experiment at the LHC accelerator at CERN relies on its computing
infrastructure to stay at the frontier of High Energy Physics, searching for
new phenomena and making discoveries. Even though computing plays a significant
role in physics analysis we rarely use its data to predict the system behavior
itself. A basic information about computing resources, user activities and site
utilization can be really useful for improving the throughput of the system and its
management.  In this paper, we discuss a first CMS analysis of dataset popularity
based on CMS meta-data which can be used as a model for dynamic data placement
and provide the foundation of data-driven approach for the CMS computing infrastructure.
\end{abstract}

\section{Introduction}

The Large Hadron Collider at Geneva, Switzerland, was designed to uncover
mysterious parts of Universe building blocks.
The tremendous efforts of physicists all around the globe were rewarded by
Higgs discovery through analysis of Run-I data from both ATLAS and CMS
experiments~\cite{Higgs1,Higgs2}. The great success of Run-I was strongly
dependent on reliable computing infrastructure. In CMS, the computing model
is quite complex. The data collected from the detector are streamed to the HLT
farm and organized into trigger streams. Later they are archived at the Tier-0
center at CERN and distributed to CMS Analysis facility (CAF) at CERN
and to Tier-1 centers around the globe. Many Tier-2 centers worldwide
share a portion of the data for further processing and Monte-Carlo generation.
Finally, the Tier-3 centers (mostly at University levels) are used for various
analysis tasks. Such distributed Computing Model is demonstrated to be
reliable and flexible enough during Run-I operations, see~\cite{CMSComp}.

During Run-I, CMS collected and processed more than 10B data events and
simulated more than 15B events. The data transfer throughput peaks at few PB
per week placing the data across hundreds of GRID sites for various user activities,
ranging from data operations tasks to individual physics analysis.  Even though CMS is
able to successfully handle the data through Run-I operations
\cite{CMSDataManagement} the experiment is looking at different ways to improve the system
and reduce its operation cost and manual labor.

In this paper, we present a first attempt to understand different
aspects of the CMS dataset popularity by the analysis of the CMS computing model.
We start with the problem statement in Section~\ref{Sec2},
followed by discussions of various streams of information at our
disposal from CMS data management systems.
Section~\ref{Sec3} provides a description of the architecture and implementation of different
components used by Machine Learning (ML) algorithms to model
and predict CMS dataset popularity. In particular, we discuss 
time series (referred to as ``rolling'') analysis for future
predictions, operational cost of results obtained by ML algorithms
and seasonality aspect of the data.

\section{Problem statement\label{Sec2}}
The CMS dataset popularity is a metric that quantifies user activities in
CMS \cite{DDPPaper}. The data produced by the detector are logically organized into
hierarchical structure, such as runs, files, blocks and datasets. The data
coming from the detector are grouped into runs which represent a unit of
the data-taking process. These runs are stored into files which are suitable for
data processing by reconstruction or analysis programs.  The files themselves
are grouped into blocks which are used by data operations teams as atomic units for
data transferring among the sites. Blocks are grouped into datasets which
represent a processing chain of specific physics processes. Once data is produced
and become available for end-users they use datasets as a common unit to submit
their processing or analysis jobs.  Every job will resolve a given dataset into
series of files and process them at once. The outcome will be either stored in
other files and a new dataset will be formed or it can be converted into n-tuples,
histograms or any other form used by physicists in their analysis.  Therefore
it is possible to measure user activities in terms of dataset popularity, i.e.
those datasets which are used in analysis jobs more often than others can be
considered as popular.

During Run-I we collected more than 1TB of meta-data stored in various
databases.  More than 400K datasets, 4.5M blocks and 100M files were recorded
in data bookkeeping system (DBS), see \cite{CMSDataManagement}. Only a portion
of these datasets were actively accessed by end-users. The other parts were
transferred once and had a very low access rate. The latter can lead to sub-optimal
utilization of CMS storage space as well as overhead in the data management
system with unnecessary data transfers. On the other hand, the
reduction of dataset access latencies can be very useful for load balancing and
scheduling jobs among the sites. This becomes especially important during rush
periods when results are prepared for major events such as conferences.
Therefore it is desirable to understand the nature of CMS dataset popularity in
order to improve our computing model and reduce its operational cost.  Among
different possibilities to tackle these issues we decided to model and predict
CMS dataset popularity via ML algorithms based on historical usage of
datasets via various meta-data information available to us.

The CMS data management system is composed by several systems
\cite{CMSDataManagement, DDPPaper}: PhEDEx (dataset replication system), DBS (data
bookkeeping catalog), Dashboard (global monitoring system which keeps tracks of
user jobs), SiteDB (site and user database), PopDB (CMS datasets popularity
database), DDM (Dynamic Data Management system) and others\footnote{Only the
systems which are relevant to and involved in the CMS datasets popularity are listed here.}.
The PopDB database keeps historical information of datasets usage in
various metrics, e.g. number of total and user accesses per dataset, as well as
total CPU time spent on each datasets during processing or analysis jobs. This
information is used by DDM to replicate popular datasets to other sites.
But the latter happens post-factum and requires sufficient amount of historical
data to be accumulated up-front to trigger the replication process. To fill
this gap we proposed to investigate a possibility to predict CMS dataset
popularity based on existing meta-data for newly created datasets and use these
predictions as a seeding mechanism for DDM machinery.

\section{Architecture \& Implementation\label{Sec3}}
We started with evaluation of CMS data services and their underlying
back-ends as a possible sources of information for CMS dataset popularity problem.
We divided them into three categories: structured data which can be obtained
from CMS data-services such as DBS, PhEDEx, Dashboard, SiteDB, PopDB; the
semi-structured data which can be found from CMS web logs, calendar systems
and un-structured data, such as HyperNews forums, CMS twikies, etc.
Obviously obtaining information from structured and semi-structured sources is
much easier than un-structured ones. The latter will require significant
pre-processing effort which is far beyond our abilities at the
moment\footnote{In fact, we foresee the usage of un-structured data but such
effort should be justified at later stage of the project.}. Therefore, we
concentrate our work only on first source of information, i.e.
structured data extracted from CMS data-services.

\begin{figure}[htb]
\begin{center}
\includegraphics[width=.7\textwidth]{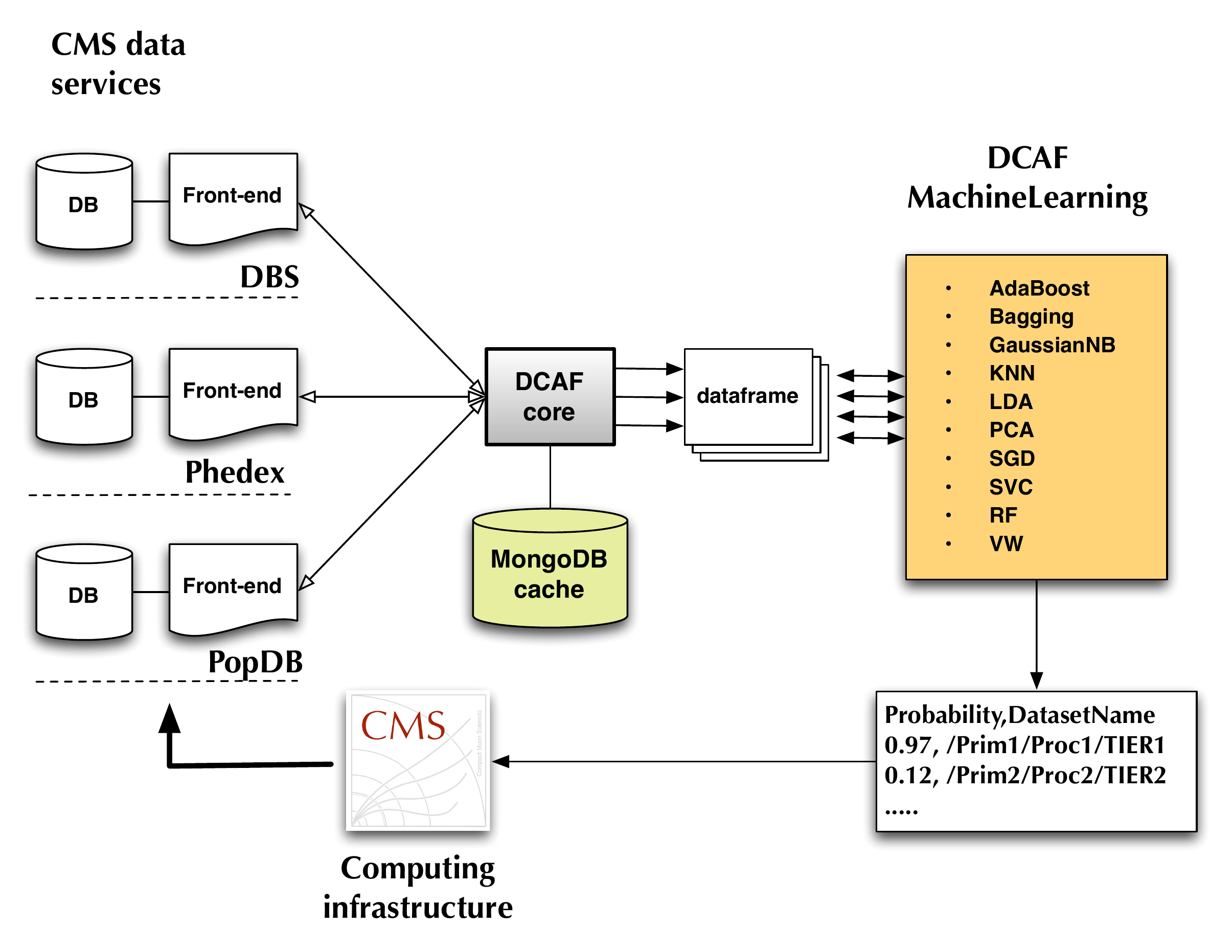}
\caption{Flow diagram of DCAFPilot framework which collects and pre-processes information from
various CMS data-services, anonymizes information within internal cache and passes them on to underlying
ML algorithms.}
\label{FIG:DCAFPilot}
\end{center}
\end{figure}

We developed DCAFPilot (Data and Computing Analysis Framework Pilot)~\cite{DCAFPilot}
framework in order to understand the metrics, the analysis
workflow and the necessary tools (with possible technology choices) needed to
tackle this problem.  Its overall architecture is shown in Figure
\ref{FIG:DCAFPilot}.  It collects information from various CMS data-services,
passes it to a pre-processing step which ensures that information stays anonymous
and later to external ML algorithms for model building. The predictions are
represented in a form of probabilities and dataset names which later can be
fed back into the CMS computing infrastructure, e.g.
used by DDM system as initial seeds for dataset replication.

The DCAFPilot framework was implemented in python and integrated with
the scikit-learn ML library \cite{scikit}. We also supplement it by 
Yahoo Vowal Wabbit \cite{VW} and Distributed Gradient Boosting xgboost
ML libraries \cite{xgboost}. As a baseline we used five algorithms: RandomForest \cite{RF},
LinearSVC \cite{SVC}, SGDClassifier \cite{SGD}, VW \cite{VW}, and xgboost \cite{xgboost}
to establish automated machinery to
build and train the model, predict CMS dataset popularity in up-coming week and
understand the cost of the modeling procedure.

\subsection{Data collection\label{Sec3.1}}
We used five CMS data-services (DBS, PhEDEx, SiteDB, Dashboard, PopDB) to
collect all required meta-data information suitable for ML analysis.  In
addition, the DBS services consists of four different instances used to publish
globally accessible datasets and physics analysis ones and we used all of
them during data collection phase. The data from CMS services were available via
data-service APIs.  In total, we utilized about a dozen of APIs from
aforementioned CMS data-services, places 800K queries about 200K datasets, 900
software releases, 500 site entries and 5K user entries. The obtained data were stored in
an internal cache, see Figure \ref{FIG:DCAFPilot}, and anonymized either by using
internal database ids or via internal hash function. The data collection
procedure was automated via UNIX cronjobs.  The data were organized by weeks
with total of 52 data files per year, where each file has 82 attributes.
The datasets were extracted from PopDB and complemented by datasets from the DBS
system. The former datasets were used to define dataset popularity metric, while latter ones represented
``unpopular'' data. One year of data was roughly translated into the 82$\times$600000
dataframe\footnote{Here and further we explicitly distinguish a CMS dataset vs
a dataset suitable for ML algorithm which we call a dataframe.}.  In total,
we collected 130 dataframe files (52 files per year), worth of $\sim1.5$M
measurements for different CMS datasets spawning over the years 2013-2015.
Separately, we also generated conference count dataframe based on the CMS CINCO
\cite{CINCO} database. It contains counters for up-coming conferences where
CMS presents its results. The counters collect total number of conferences
in 1,2,4,6,10 weeks followed by 5 week intervals up-to 70 weeks in the
future. This dataframe was used as a complement to dataset meta-data
to study the influence of conferences on the
CMS dataset popularity (see Section \ref{Sec3.6} for discussion).

The collected data presents a good representation of various user activities
through 2013-2015 years. These years cover Run-I era, preparation for Run-II
and future detector upgrade studies. And, we expects that such activities will
only grow through the Run-II phase and beyond.

\subsection{Dataset popularity\label{Sec3.2}}
Our first task was to define dataset popularity metric.  As simple as it sounds it
had many dimensions.  The popularity DB provides us six different metrics: the
number of accesses to a given dataset, the number of users/day recorded for a dataset,
the total number of CPU hours spent on a given dataset, along with normalized
values of those attributes over full number of datasets.

\begin{figure}[htb]
\includegraphics[width=.5\textwidth]{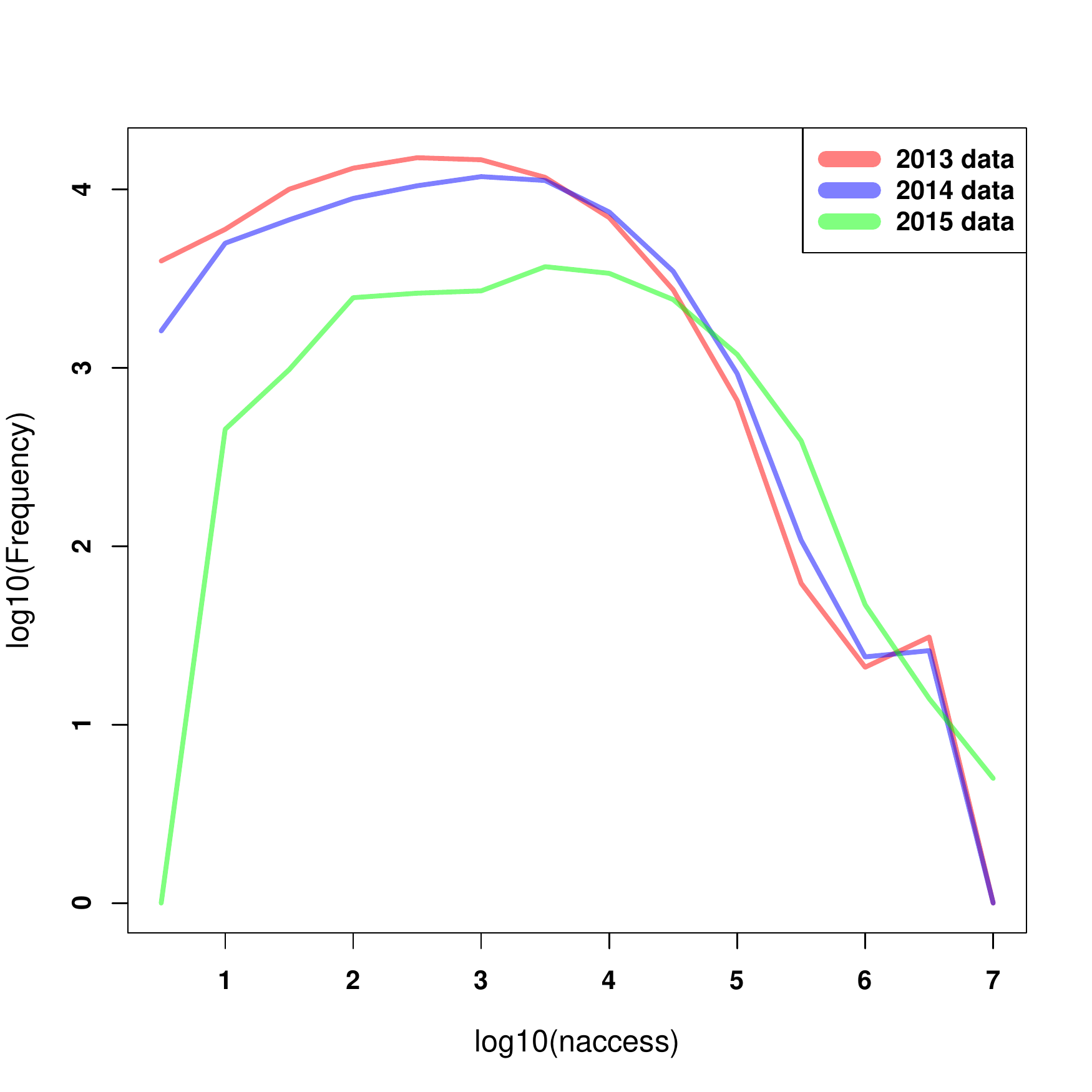}
\includegraphics[width=.5\textwidth]{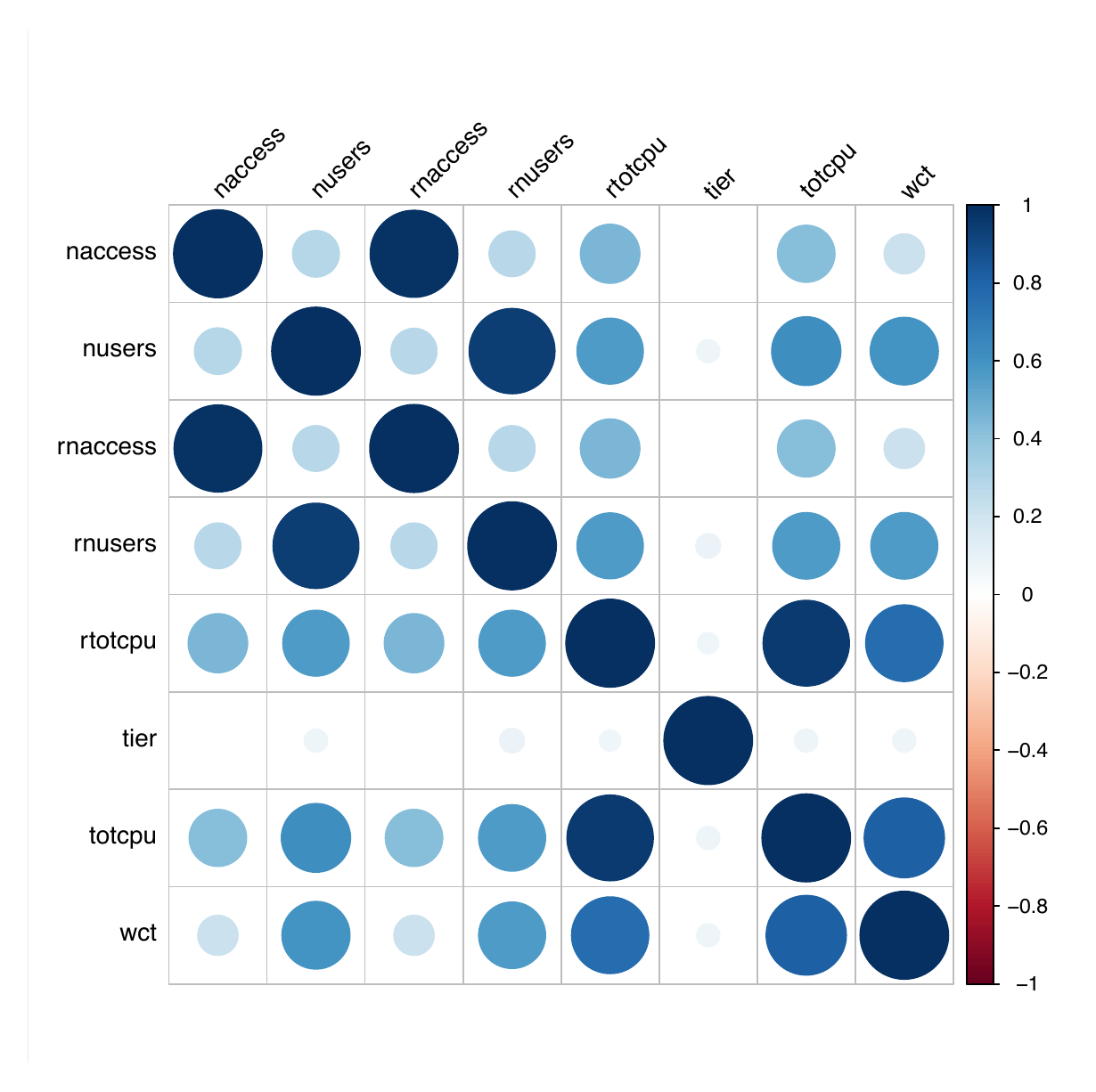}
\caption{
Number of accesses per dataset and sub-set of
available attributes in a correlation plot for the 2013-2015 data sets.
}
\label{FIG:PopularityMetrics}
\end{figure}

Figure \ref{FIG:PopularityMetrics} shows the behavior of some metrics recorded
in popularity DB. Even though the shape of distributions of different
years looks alike,
we found that data themselves required further clean-up and understanding.
For instance, number of users at low access patterns had gaps which
represent the fact that our users were picky about certain datasets. The overall
value of different metrics in 2015 were lower with respect to 2013/2014 years
which reflects the usage of 2015 data sample (we only used half of the year
available at the time of our studies). And, we have significant correlations
among some attributes, e.g. number of dataset accesses is highly correlated
with number of users used those datasets in their jobs, see Figure
\ref{FIG:PopularityMetrics}.

In addition, we were dealing with datasets composed out of 59
different data tiers within our sample. Each data-tier represents a specific
interests within physics community, e.g.  the raw data, processing data, the
analysis data etc. The further work was only concentrated on five different
data-tiers: AOD - Analysis Object Data, AODSIM - AOD with simulated
information, MINIAOD - reduced AOD objects, MINIAODSIM - MINIAOD with simulated
information and USER data-tier to denote user-based datasets. These five
data-tiers represent the main interests for dataset popularity since they
correspond mostly to main user activities, such as analysis and data-processing,
rather than daily activities of various data operations teams within the CMS
experiment. The latter can be considered as ``routine'' tasks and obviously are
well-defined within the experiment.

\begin{table}[htb]
\begin{center}
\caption{Effects of applied cuts on the different data tiers.}\label{TAB:cuts}
\begin{tabular}{lllllll}
    \hline
    \hline
Data tier/Cut& No cut        & naccess$>$10 & log(nuser)$>$2 \\
    \hline
AOD          & 4924 (7.25\%) & 4687 (8\%)   & 1285 (35\%) \\
AODSIM       & 21090 (31\%)  & 18825 (32\%) & 1547 (42\%) \\
MINIAOD      & 7 (0.01\%)    & 6 (0.01\%)   & 0           \\
MINIAODSIM   & 1083 (1.5\%)  & 792 (1.3\%)  & 28 (0.8\%)  \\
USER         & 34127 (50\%)  & 28777 (49\%) & 380 (10\%)  \\
    \hline
ALL          & 67892         & 59222        & 3683        \\
    \hline
    \hline
\end{tabular}
\end{center}
\par
\bigskip
\end{table}

\begin{figure}[htb]
\includegraphics[width=.55\textwidth]{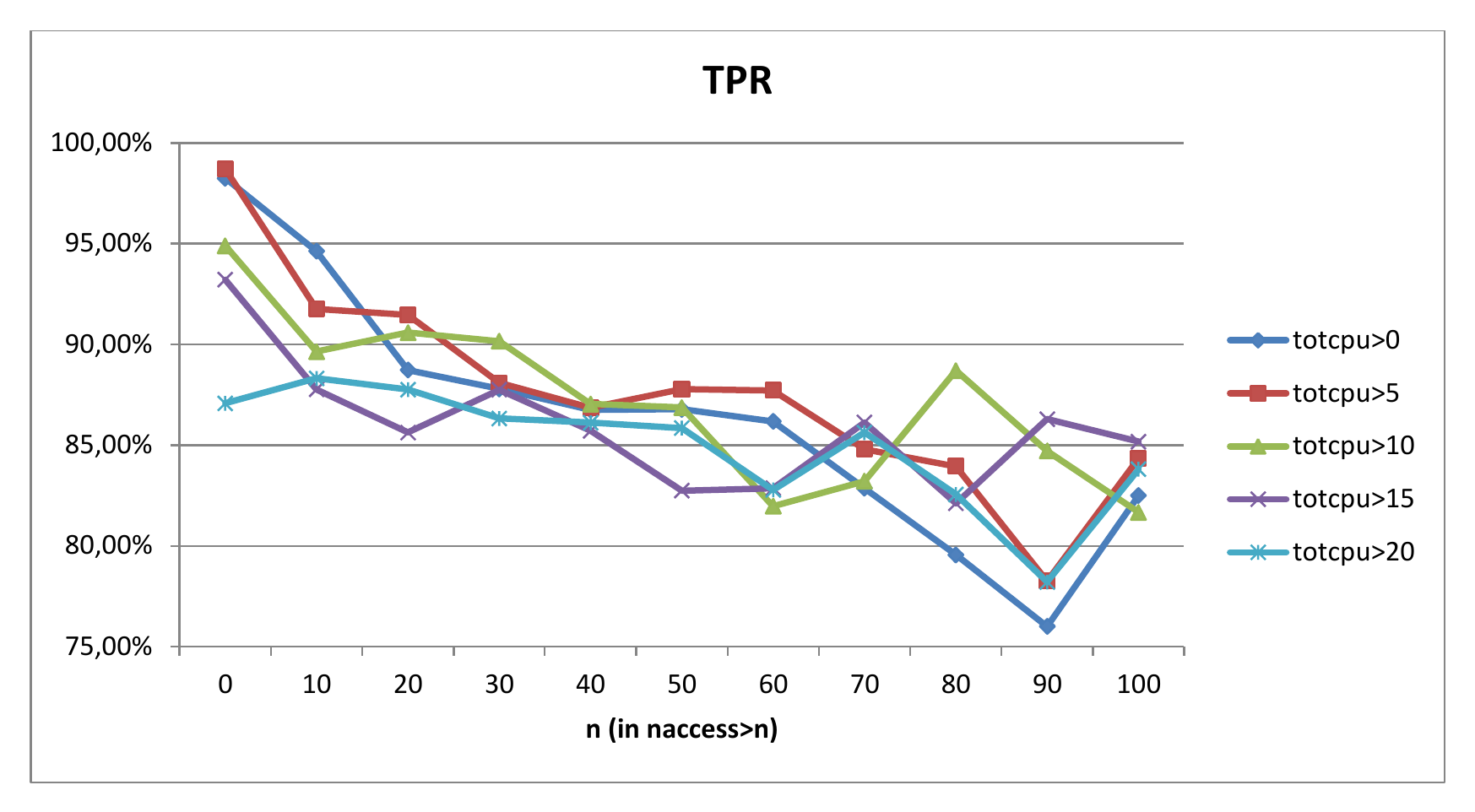}
\includegraphics[width=.5\textwidth]{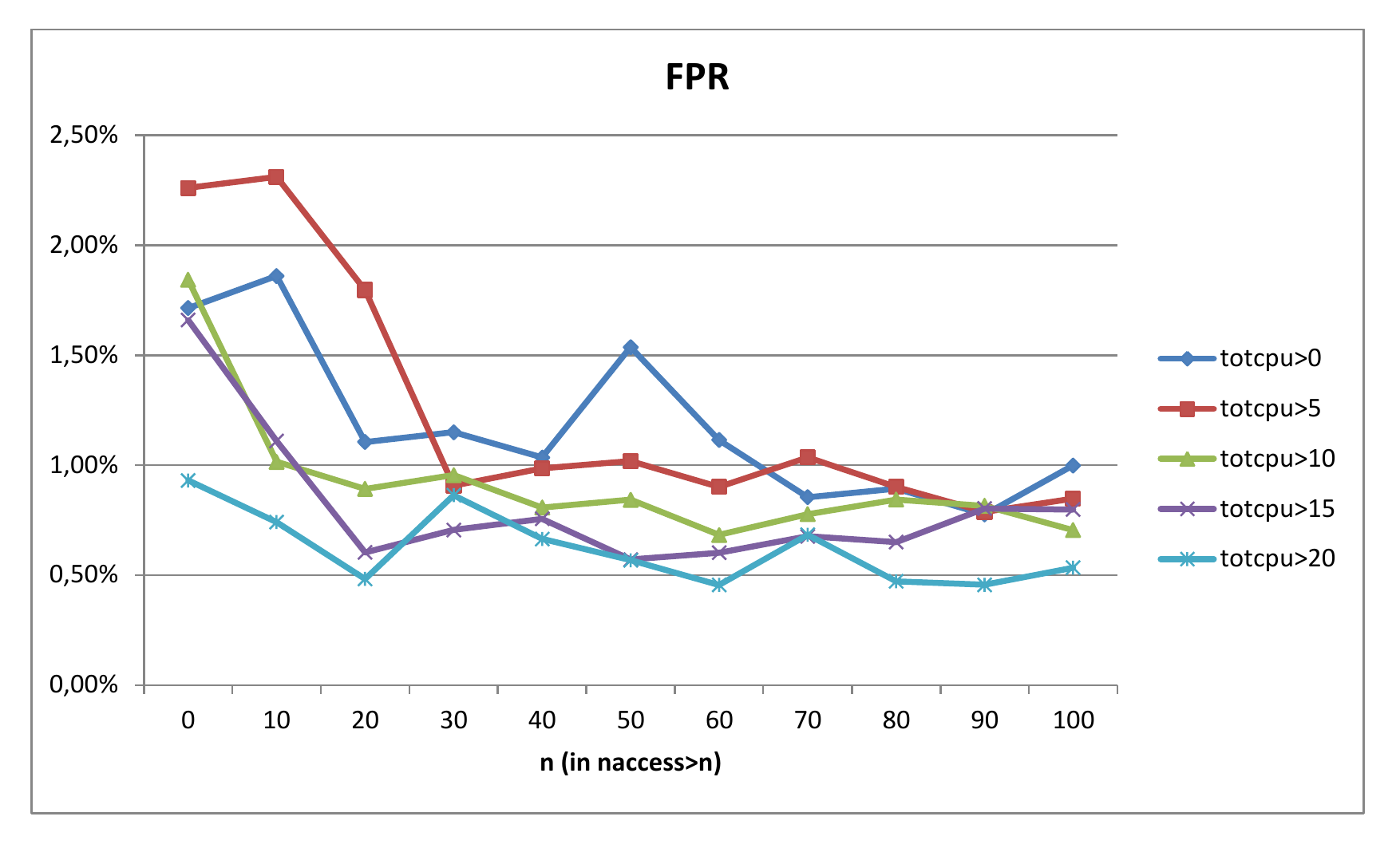}
\caption{
    True positive and False positive rates for different
    values of $naccess$ and $totcpu$ parameters using Random Forest
    classifier over 2014 datasets.
}
\label{FIG:TPR_FPR}
\end{figure}

We collected the data for 2013, 2014 and first half of 2015 data and analyzed
data-tiers breakdown within these samples. Table \ref{TAB:cuts} shows the
total number of datasets recorded in PopDB along with cuts for different
data-tiers within 2014 dataframe. Further, we performed analysis of
various cuts based on yield of TP (true positive),
FP (false positive), TN (true negative) and FN (false negative) rates
using Random Forest classifier.
The Figure \ref{FIG:TPR_FPR} shows model behavior under different $naccess$ and
$totcpu$ cuts. Based on this studies we found that $naccess>10$ and $totcpu>10$
or $naccess>10$ cuts provide the most the most promising TP/FP/TN/FN rates. For further
studies we opted for simple threshold, $naccess>10$.

\subsection{Workflow\label{Sec3.3}}
Our workflow was organized in the following way. The data was collected on a
weekly basis via periodic cronjob which requested dataset information from
PopDB for a given week. For every dataset in this set we
obtained meta-data information from DBS, PhEDEx, SiteDB and Dashboard systems. This
data was complemented by using random datasets from DBS system.
The average time to collect 1 week of data took several hours and fade out
gradually upon filling out internal cache. It was possible to run these jobs
within single VM without requiring for additional resources.
As mentioned in Section \ref{Sec3.1} we added to each
dataframe the ``unpopular'' datasets meta-data with 1:10 ratio. 
This ratio was used to supplement signal data from PopDB with noise one
from DBS system, i.e. we used random datasets from DBS which had no records
in PopDB for given time period. All data attributes were
anonymized via internal cache hashing mechanism and checked for correlations.
Even though we were able to get 82 attributes representing each dataset we
dropped 12 attributes due to their bias towards the popularity
metric\footnote{For example, number of sites which hosted a given dataset
certainly is biased towards popularity predictions since it represents posterior information.}.
We used additional transformation step to convert obtained dataframes into
classification problem, for which we used a cut discussed in Section
\ref{Sec3.2}. At this step the data were ready to be fed into any ML library.

\subsection{Processing\label{Sec3.4}}
Once the data were transformed into suitable format for ML we used
the ``rolling'' approach which consists of the following steps:
\begin{itemize}
    \item obtain a set of data up to the week in question (usually we kept one year worth of data) and train a specific classifier;
    \item obtain new data and make a prediction for selected data-tiers (we are only interested in
        user based data and AOD datasets);
    \item use the prediction as a seeding for DDM service who allocates replicas of popular dataset at various sites;
    \item once data for the week in question becomes available in PopDB, compare
        the predictions to the actual dataset access patterns.
        This step can be used to adjust the model in the future.
\end{itemize}

The former two steps were performed with five ML classifiers: RandomForest,
LinearSVC, SGDClassifier, VW and xgboost (see Section \ref{Sec3}), and standard
70/30 split was used to test the model via cross-validation.  We measured the
outcome of True Positive (TP), True Negative (TN) rates for different
data-tiers and estimated the operational cost of False Positive (FP) and False
Negative (FN) rates.
In the following, the results obtained from these studies are presented; the
last two steps in the previous list are under implementation in the production
system.

\subsection{Classification results\label{Sec3.5}}
We used 2013 data sample as the initial dataframe
and gradually added weekly data to measure TP/TN/FP/FN rates for data through 2014.
Among all used classifiers we found that SGDClassifier
performs best, giving less than 10\% error for TP/TN rates for the majority of the weeks.

From the results of various classifiers we
found that prediction errors are larger during ``vacation'' periods, i.e.
first and last weeks of the year, when CMS users are less active. During
normal operation during the year all classifiers shown that it was feasible to keep TP/TN error
below 10-15\% using their default parameters. The fine tuning
of algorithms were not performed since we mostly concentrated
on setting up machinery for our task. The large deviations of errors
were caused by different sensitivity of underlying algorithm to
specific data tiers. Some data tiers, e.g. MINIAOD,
were introduced during 2014 and did not gain too much
access pattern among our users and therefore had larger errors. While others, e.g.
AOD, AODSIM, and USER, had smaller errors throughout the entire year. Table \ref{TAB:rates}
summarizes the mean and standard deviations for different statistics metrics
grouped by data-tiers.

We found that only AOD, AODSIM and USER data-tiers gave reasonable predictions
and their errors are stable through the entire year. The MINIAOD and MINIAODSIM
datasets were less sensitive to algorithm due to the fact that they
only become available for end-user through 2014 and did not attract user interest.
In fact, they represented only 1.5\% of entire data sample used in this analysis.

\begin{table}[htb]
    \caption{
Summary of TPR (true positive rate, i.e. sensitivity),
TNR (true negative rate, i.e. specificity),
FP (false positive) and FN (false negative) values
for different data tiers.
    }\label{TAB:rates}
\begin{center}
\begin{tabular}{llllll}
    \hline
    \hline
    Data tier  & TPR & TNR & FP & FN \\
    \hline
    AOD        & 0.97$\pm$0.05 & 0.99$\pm$0.02 & 0.005$\pm$0.011 & 0.015$\pm$0.029 \\
    AODSIM     & 0.93$\pm$0.13 & 0.99$\pm$0.02 & 0.008$\pm$0.016 & 0.021$\pm$0.045 \\
    MINIAOD    & 0.11$\pm$0.32 & 0.99$\pm$0.02 & 0.014$\pm$0.026 & 0.001$\pm$0.007 \\
    MINIAODSIM & 0.49$\pm$0.48 & 0.99$\pm$0.02 & 0.009$\pm$0.016 & 0.007$\pm$0.031 \\
    USER       & 0.93$\pm$0.15 & 0.98$\pm$0.02 & 0.014$\pm$0.021 & 0.011$\pm$0.023 \\
    \hline
    \hline
\end{tabular}
\end{center}
\end{table}

The obtained errors on FP rates can be easily translated into data-transfer and
extra disk space overhead we may face. For example, an average size of CMS
dataset is around 2TB. If we have 1\% of FP rate, see Table \ref{TAB:rates},
and take on average 500 datasets created in CMS\footnote{We looked at
distribution of newly created datasets through 2014 and found that it has
$565\pm300$ datasets on average.} this will translate into extra 10TB of
data-transfer per week out of few PB of data transfer CMS is doing on weekly
basis.  Therefore the overhead would be $\sim1\%$ of total transfer rate. Such
overhead can be easily accounted by data transfer team.  We found that the cost
of FN rate can be considered as ``irrelevant'' from operational point of view
since it will correspond to current mode of operations in CMS, since data
placement is done manually via current snapshot of site occupancy.  But in the ideal
world where computing resources will be driven by such data popularity
predictions the FN rate will translate into longer latencies of
analysis user jobs.
On average we have 1.5M jobs submitted every day\footnote{We
used Dashboard CMS data-services and found that on average CMS has
$1520669\pm450069$ completed jobs per week.}. The 1\% of FN rate will
correspond that $\sim15$K jobs will compete for resources of a site where
popular dataset may resides.
On another hand, the data driven approach would
reduced overhead of site utilization. So far, our sites are
populated with data which we put over there rather the data which will
become popular. For example, we found that majority of datasets
are accessed only by a few users.
Therefore a better data placement will directly translate into site utilization
savings, further studies of which are underway.

\subsection{Prediction from information beyond dataset features\label{Sec3.6}}
Even though we used historical information to predict popularity
of datasets in a future we also considered that future events, such as
up-coming conferences, group meetings and discussion within
various physics groups can stimulate access patterns to certain
datasets. Such studies will require data-mining of user based discussion
groups, meetings agendas and up-comping conferences. As a starting
point we used the CINCO CMS conference database \cite{CINCO} where physics conferences
are registered and physicists are assigned to present their
work on physics analysis.

Naively, we thought that some datasets which represent certain physics channels
may become popular before up-coming conferences where physicist present their
results. To test this idea we evaluated if the future conference weekly counts - if used
together with the features of the datasets - could improve the classification
of whether a dataset is popular or unpopular, with respect to using only the
features of the dataset.
This was done by comparing the
classification accuracies, comparing the feature importance scores given by
random forest classifiers for individual features, and comparing the p-values
of the $\chi^2$-independence test and the ANOVA F test between individual
features and the popularity outcome. The experiment has not shown improvement
yet, in part because the classification performance was already nearly perfect
for the limited time period when the conference data was available.  This also
lead us to explore the influence of the conference data from alternative
perspectives.

\begin{figure}[htb]
\begin{center}
\includegraphics[width=.6\textwidth]{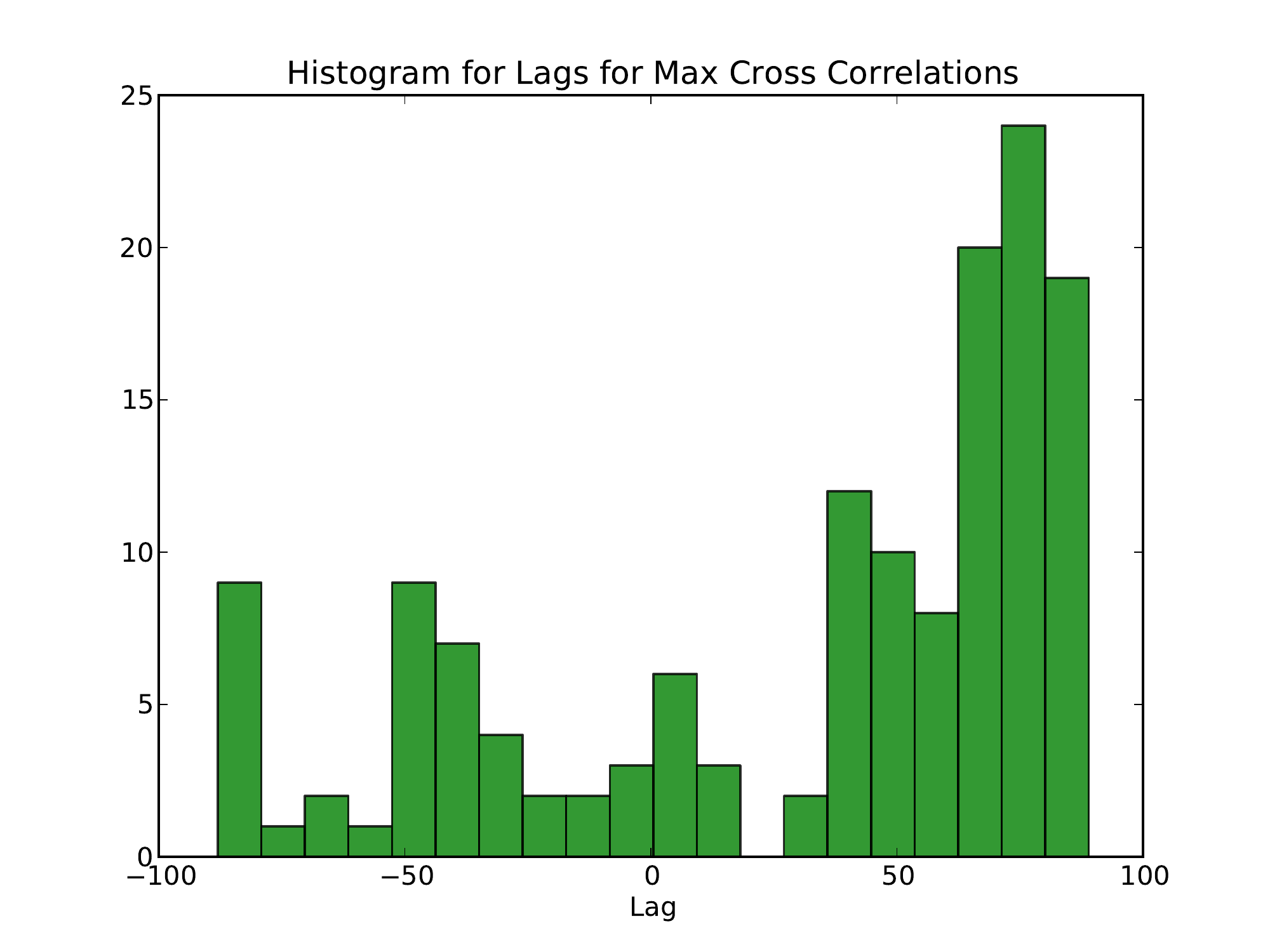}
\caption{Histograms of lags with maximum cross correlation (see text for more details).}
\label{FIG:hist}
\end{center}
\end{figure}

For instance, we calculated the cross correlation between the weekly count time
series of conferences and the one of the accesses to each dataset at different
time lags, and found that the time lag at which the cross correlation peaks is
roughly 75 weeks for most datasets.  We also found that for some datasets cross
correlation is peaking at a positive lag implying that future conference
schedules can affect the current dataset access, while for others it peaks at a
negative lag meaning that past conferences can still have residual influence on
the current dataset access. Figure \ref{FIG:hist} shows that such lags have significant cross correlations.

\begin{figure}[htb]
\includegraphics[width=.55\textwidth]{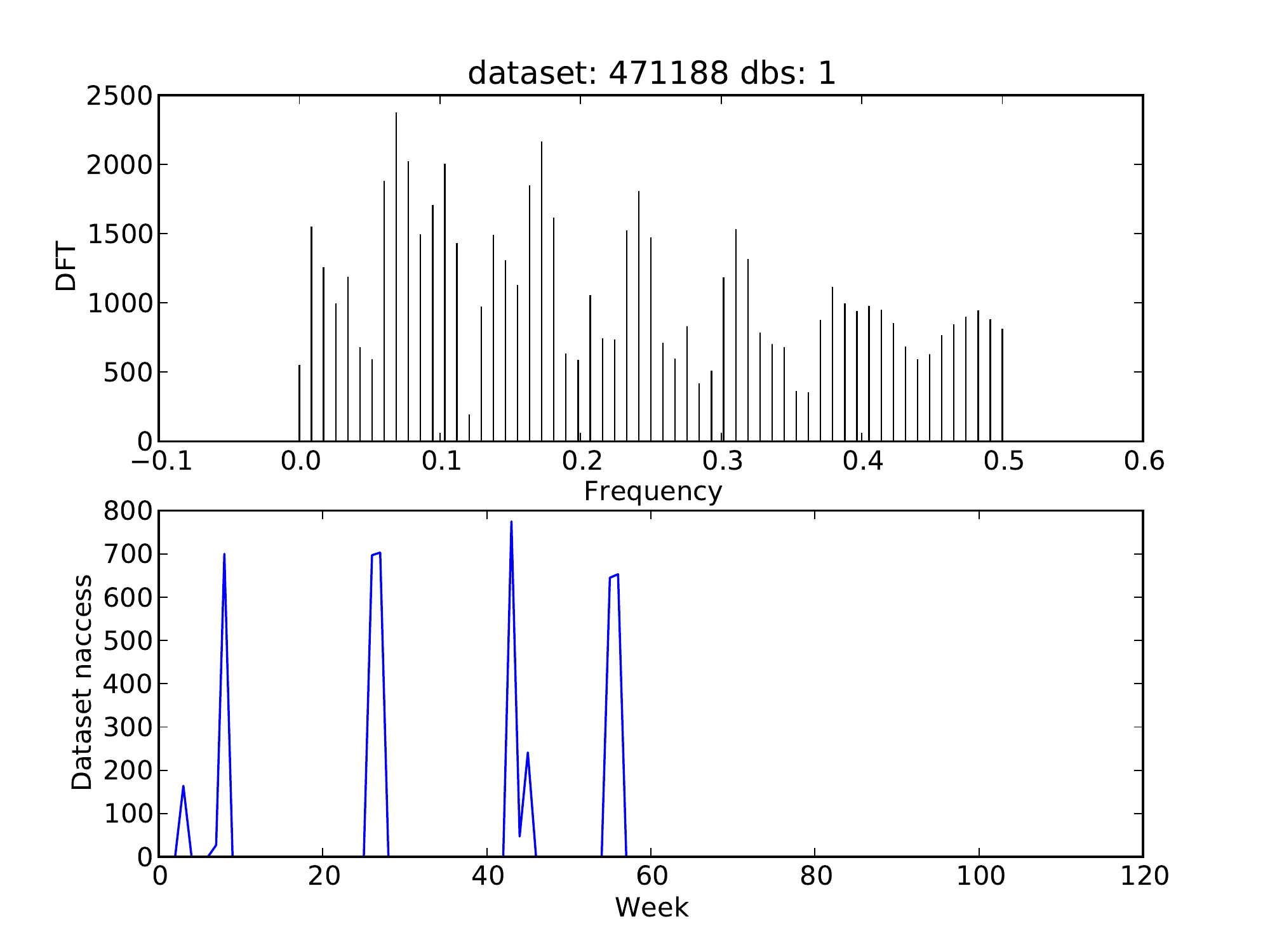}
\includegraphics[width=.55\textwidth]{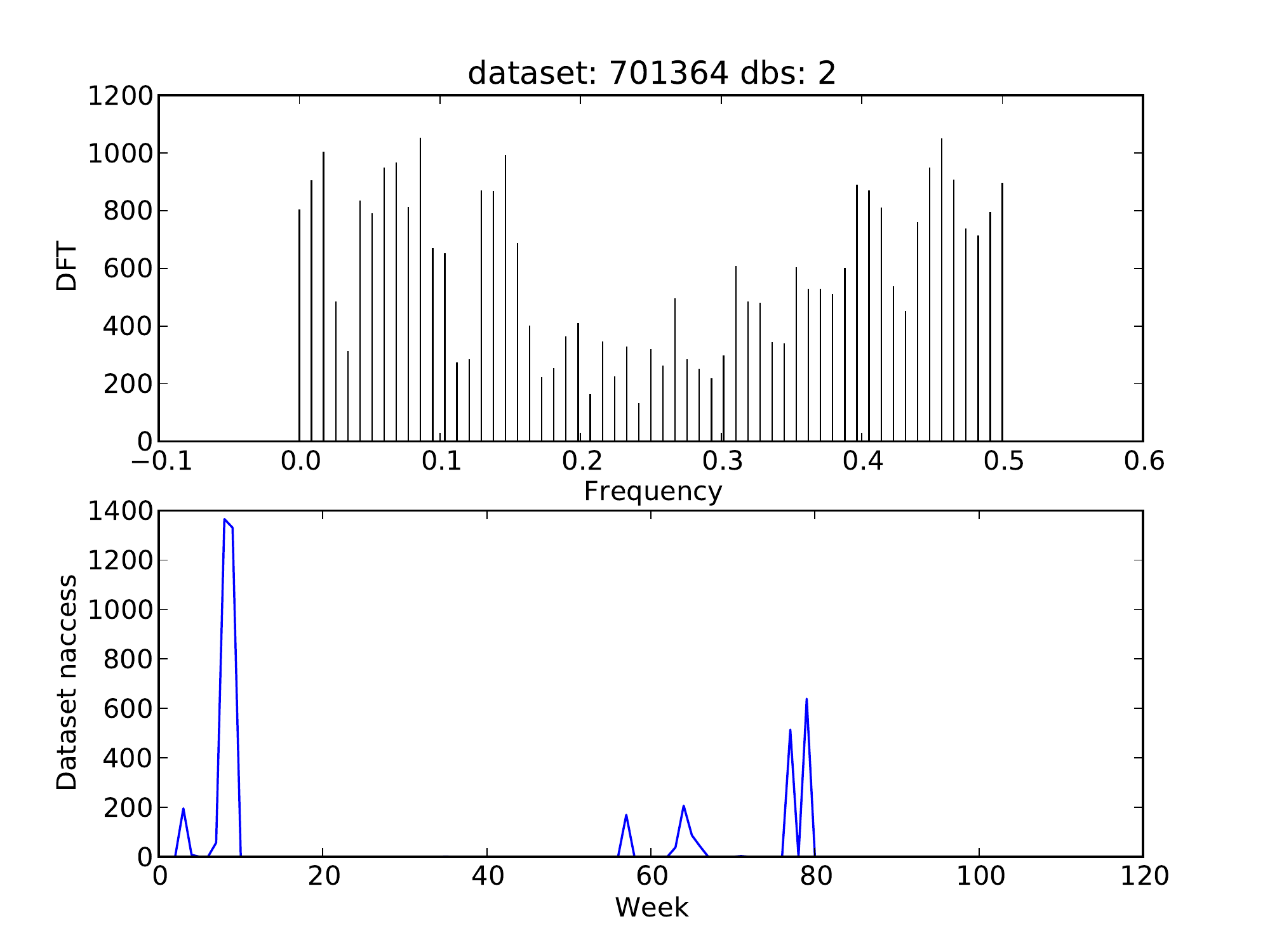}
\caption{Seasonality effect of random datasets from 2014 data sample.
Both plots show DFT of access count per week time series of a dataset.
The length of time series is 116 weeks.
The highest spike of top plot for dataset ID=471188 occurs at the 8-th frequency and
corresponds to a time period of 116/8 = 15 weeks. The DFT plot
of another dataset ID=701364 does not show seasonality in the dataset
access series.}
\label{FIG:seasonality}
\end{figure}

Finally, we tried to find out if each time series contains significant
seasonality, i.e. periodicity. This was done by calculating the Discrete
Fourier Transform (DFT) on the time series using the Fast-Fourier Transform
(FFT) algorithm and searched for the most significant spike that represents the
frequency of seasonality. We found a yearly seasonality effect for the
conferences, a shorter-period seasonality for some datasets (for example,
15-week period), and no seasonality for other datasets (see Figure \ref{FIG:seasonality}).

These studies demonstrate that the conference count and some CMS datasets
access counts can show some seasonality, and there is significant correlation
between conference count and most dataset access counts. But further studies
are required to see if such observations can improve the overall prediction for
dataset popularity, and it is especially interesting in the context of specific
data-tiers.

\section{Summary}
The results of this pilot project provides a solid foundation and a working
pilot framework for future R\&D work on data mining of several aspects of the
CMS computing model. We demonstrated that we are able to successfully predict
CMS dataset popularity based on collected CMS meta-data.  The obtained results
shows that small FP/FN rates are possible to achieve and can be interpreted as
overhead in data-transfer and job latencies. We estimated that obtained FP rate
can introduce only 1\% overhead to existing data transfer rate and can be
easily managed by operational team. The small FN rate can lead to better sites
utilization and provide operational savings in the long term.  We are
evaluating to turn this pilot project into production and use dataset
popularity predictions as initial seeding for the DDM \cite{DDPPaper} system.
In addition, we are interested in further investigations of seasonality effects
for specific data tiers. The relatively small errors for the few discussed
data-tiers can be turned into better data placement and reduced latencies of
analysis jobs during first week of data appearance. This will lead to better
analysis throughput and user satisfaction with their data processing needs,
more awareness in dynamic data placement tactics across different sites, as
well as better usage of site resources in the CMS GRID infrastructure.

\section*{Acknowledgment}
This work was partially supported by the National Science Foundation, contract No.
PHY-1120138, and the Department of Energy of the United States of America.


\section*{References}
\begin{thebibliography}{9}

\bibitem{Higgs1} ATLAS experiment,
{\it Observation of a new particle in the search for the Standard Model Higgs boson with the ATLAS detector at the LHC},
Physics Letters B, Volume 716, Issue 1, pp 1-29, 2012.
\bibitem{Higgs2} CMS experiment,
{\it Observation of a new boson at a mass of 125 GeV with the CMS experiment at the LHC},
Physics Letters B, Volume 716, Issue 1, pp 30-61, 2012.
\bibitem{CMSComp}  J. Adelman et al, {\it CMS Computing Operations during Run 1},
Journal of Physics: Conference Series 513 (2014) 032040.
\bibitem{CMSDataManagement}
M. Giffels, Y. Guo, V. Kuznetsov, N. Magini and T. Wildish
{\it The CMS Data Management System}
J. Phys.: Conf. Ser. 513 042052 doi:10.1088/1742-6596/513/4/042052 
\bibitem{DDPPaper} F.H. Barrerio Megino et al,
{\it Implementing data placement strategies for the CMS experiment based on a popularity model}
J. Phys.: Conf. Ser. 396 032047 doi:10.1088/1742-6596/396/3/032047
\bibitem{ISGC} V. Kuznetsov, T. Wildish, L. Giommi, D. Bonacorsi
{\it Exploring patterns and correlations in CMS Computing operations data with Big Data analytics techniques}
International Symposium on Grids and Clouds (ISGC) 2015, Academia Sinica, Taipei, Taiwan
\bibitem{DCAFPilot} https://github.com/dmwm/DMWMAnalytics/
\bibitem{scikit} http://scikit-learn.org/stable/
\bibitem{VW} https://github.com/JohnLangford/vowpal\_wabbit/wiki
\bibitem{xgboost} https://github.com/dmlc/xgboost
\bibitem{RF} L. Breiman, {\it Random Forests}, Machine Learning 45 (1) 5–32. doi:10.1023/A:1010933404324
\bibitem{SVC} C. Cortes, V. Vapnik, {\it Support-vector networks}, Machine Learning, 20, 273-297 (1995),
\bibitem{SGD} {\it Stochastic Gradient Descent}, Website, 2010.
    http://leon.bottou.org/projects/sgd
\bibitem{CINCO} Cms INformation on COnferences (CINCO) data-service, https://cms-mgt-conferences.web.cern.ch/
\end{thebibliography}
\end{document}